\documentclass{aa}
\usepackage{graphicx}

\begin{document}

\title{The Star Cluster population of NGC 5253
      \thanks{Based on observations made with the ESO VLT,
       program 69.B-0345, and on HST observations obtained from
       the ESO/ST-ECF Science Archive Facility} 
}

\author{}
\author{G. Cresci \inst{1} \and L. Vanzi \inst{2} \and M. Sauvage \inst{3}}
\institute{}
\institute{Dipartimento di Astronomia, Universit\'a di Firenze,
    Largo E. Fermi 5, I-50125, Firenze, Italy
        \and ESO, Alonso de Cordova 3107,
    Casilla 19, Santiago 19001, Chile
        \and Service d'Astrophysique, CEA/DSM/DAPNIA,
    Centre d'\'Etudes de Saclay, F-91191 Gif-sur-Yvette Cedex, France}

\offprints{G. Cresci \\
    \email{gcresci@arcetri.astro.it}}

\date{Received / Accepted}

\abstract{We present a detailed analysis of the star cluster population in
the starburst galaxy NGC~5253. Our work is based on HST optical and VLT Ks
images. We detect more than 300 clusters, and for all of them we derive the 
photometry in the V, I and Ks bands and built a color-color V-I/V-Ks diagram. 
After correction for star contamination, we matched 115 sources in all the 
three bands V, I and Ks.
Comparison of the data points with models shows that most of the sources are 
affected by a red excess. The ages of the younger clusters are obtained from 
an HST H$\alpha$ image which makes it possible to partially remove the 
degeneracy between age and extinction and to calculate the screen optical 
depth toward the clusters.
The V and Ks luminosity functions show a marked turnover at $M_V=-7.3$ and
$M_K=-10.2$, with a shape roughly similar to the Gaussian distribution of old
globular clusters. We convert the luminosity function into a mass function, 
and find that the clusters younger than 10 Myr show a power law distribution 
with slope -1.6, while the distribution of the older clusters has a maximum 
at about $4 \cdot 10^4 M_{\odot}$.
    \keywords{Galaxies: individual (\object{NGC 5253}) -- galaxies: star
        clusters -- galaxies: starburst}}

\maketitle


\section{Introduction}

NGC 5253, an irregular dwarf galaxy in the Centaurus Group, is one of the best
objects available to study young starbursts. It is in fact one of the closest
starburst galaxies, with heliocentric distance of 3.3 Mpc (Gibson et al.
\cite{gibson}), so that $1\arcsec$ corresponds to 16 parsec.
Moreover it is one of the youngest starburst galaxies known (Rieke et al.
\cite{rieke}), with an age likely $\leq 10$ Myr, as implied by the detection of
spectral features arising from Wolf-Rayet stars (Schaerer et
al. \cite{schaerer}) and from its almost entirely thermal radio spectrum with
very little synchrotron emission from supernova remnants (Beck et al.
\cite{beck}).
Its metallicity is sub-solar and about $Z_{\sun}/6$ (Kobulnicky et al.
\cite{kobulnicky}).

At centimetric wavelengths, Turner et al. (\cite{turner}) 
and Turner et al. (\cite{turner2}) 
have observed several nebulae, which they derive to be ionized by 200-1000 O stars, 
suggesting that very large clusters are the preferred mode of star formation in the 
central regions of the galaxy.

The galaxy hosts hundreds of large and massive star clusters (Caldwell et al.
\cite{caldwell}). These objects are believed to be the key features of the most intense
star forming episodes (de Grijs \cite{degrijsa}) and have been observed in many
starburst galaxies, from the extreme Luminous Infrared Galaxies (e.g. Pasquali et al.
\cite{pasquali}) to nearby dwarf irregular galaxies (e.g. Billett et al. \cite{billett},
Anders et al. \cite{anders}).
The properties of these clusters have suggested that at least a fraction of them may 
eventually be related with the population of old globular clusters observed in more 
quiescent galaxies (e.g. Boutloukos \& Lamers \cite{boutloukos}). However, their mass 
function, as measured in many young systems, does not show a preferred mass scale as 
is observed for globular clusters (Fall \& Zang \cite{fall}). This indicates that some 
evolution, for instance evaporation of non-gravitationally bound clusters, has to take 
place.
They can also be important as probes of the initial mass function in the extreme 
environments required for cluster formation, and of the formation and evolution of their 
host galaxies. \\
High spatial resolution optical observations with WFPC2 on HST by Calzetti
et al. (\cite{calzetti}, hereafter C97) have made it possible to resolve with 
unprecedented detail
the central part of NGC~5253 and to carry out a detailed study of the
brightest optical clusters and of the dust content of the galaxy.
They found that the dust reddening is markedly inhomogeneous across the
galaxy's central $20\arcsec$ region, alternating regions of small and large
reddening, so that, even in this relatively low-metallicity galaxy, the early
stage of star formation is associated with the formation of dust-embedded clusters.
For the six brightest clusters they derived ages spanning a range between 2 and
50 Myr. A mass around $10^5\ M_{\sun}$ was claimed for the four oldest clusters
(age 10$-$50 Myr), located south of the central star formation region. 
Between 20\% and 65\% of the ionizing photons in the northern star-forming 
region would be produced by the remaining two clusters. For the youngest
cluster, named cluster-5, C97 derived a mass of about $10^6\ M_{\sun}$, which makes
it a super-star cluster candidate, according to the definition by Billett et al.
(\cite{billett}).
The ages of the bright clusters in the center of the galaxy seem to be
anti-correlated with the amount of dust obscuration, supporting the idea that
clusters form deeply embedded in dust. For cluster-5, C97 derived an extinction 
$A_V>9$ mag.\\
Harris et al. (\cite{harris})  have used the same dataset of C97 to select 33 
resolved clusters detected in all the filters F300W, F547M and F814W. They have 
measured aperture photometry in all bands, extinction-corrected using the ratio 
H$\alpha$ to H$\beta$. They derived age and mass estimates for the selected clusters, 
finding that all their objects were younger than 20 Myr, with masses between a few 
thousand to $10^5\ M_{\odot}$. The young age of all the studied clusters was related 
to a particularly short dissolution timescale in NGC~5253.

Vanzi \& Sauvage (\cite{vanzi}) have observed the galaxy in the infrared and
millimeter. They used those observations together with data available in the
literature to obtain the spectral energy distribution of cluster-5. From their
analysis this cluster turned out to be a super-star cluster with a bolometric
luminosity of $1.2\ 10^{9}\ L_{\sun}$ and an equivalent O7V star number of
4700. The optical depth derived toward the cluster is $\tau \sim 7-8$.

Considering the major role of extinction in this galaxy and of dust-enshrouded
star clusters for its star formation, we add high-spatial-resolution near-IR
observations to the optical dataset used by C97, in order to sample a wider range
of optical depth and to study the properties of the extincted clusters.\\
After the description of the data set used in Sect.~\ref{obs}, we present in 
Sect.~\ref{phot} the results of the photometry of about 600 objects in the I, V and K 
band. The color-color diagram and the nebular emission are then used in 
Sect.~\ref{discuss} to derive the extinction of the clusters. 
The age of the H$\alpha$-bright clusters is computed from the equivalent width (EW) 
of the line. Finally, the luminosity and mass functions of the clusters are presented.
\begin{table*}
\caption{Observations of NGC~5253. \textit{Exp. time} is the total exposure time (for
composite images). \textit{FWHM} is the Full Width Half Maximum of a point source.}
\label{osservazioni}
    \begin{center}
    \begin{tabular}{l c c c c c}
    \hline
    \hline
    Filter & Instrument & Mean Wav. (\AA) & Bandpass (\AA) & Exp. Time (s) & FWHM \\
    \hline
    F547M & WFPC2 @ HST & 5484.0 & 483.1  & 1600 & $0\farcs17$  \\
    F656N & WFPC2 @ HST & 6564.0 & 21.4   & 3600 & $0\farcs17$  \\
    F673N & WFPC2 @ HST & 6732.2 & 21.4   & 2400 & $0\farcs17$  \\
    F814W & WFPC2 @ HST & 7921.0 & 1488.8 & 1160 & $0\farcs15$  \\
    Ks    & ISAAC @ VLT & 21600  & 2700   & 300  & $0\farcs40$  \\
    \hline
    \hline
    \end{tabular}
    \end{center}
\end{table*}

\section{Observations and data reduction} \label{obs}

In this work we have used the Ks observations of Vanzi \& Sauvage (\cite{vanzi})
combined with HST archive data.
The archival WFPC2 images (Proposal ID: 6524 and 9144) have the galaxy centered on
the WF3 chip, and they are the results of two pairs of exposures,
taken with the second pair shifted relative to the first one in order to reject hot
pixels and cosmic rays.
The F814W filter is the equivalent of the I band for WFPC2, while the F547M filter was used
for the V band. Because of the small recession velocity of the target ($404\
\textrm{Km s}^{-1}$), the F656N filter is almost centered on the H$\alpha$ line; the
F673N filter is instead used to measure the underlying stellar continuum.  
This filter contains the [SII](6712,6731) emission lines but, as in NGC5253 
the ratio H$\alpha$/[SII] is typically larger than 15 (Campbell et al. \cite{campbell}), 
we can consider this contribution as negligible for our purpose.\\
The F814W image was obtained from four exposures with 400.0, 
180.0, 400.0 and 180.0 s of integration time; the F547M image from 600.0, 200.0, 600.0
and 200.0 s exposures; the F656N image from 1500.0, 500.0, 1100.0 and 500.0 s; for
F673N only 3 images were available, with exposure times 1200.0, 600.0 and 600.0 s.
The reduction was performed using the \textit{On The Fly Reprocessing} system of the
HST archive (Swam et al. \cite{swam}).
Our data set is summarized in Table \ref{osservazioni}.\\
\begin{figure*}
    \centering
    \includegraphics{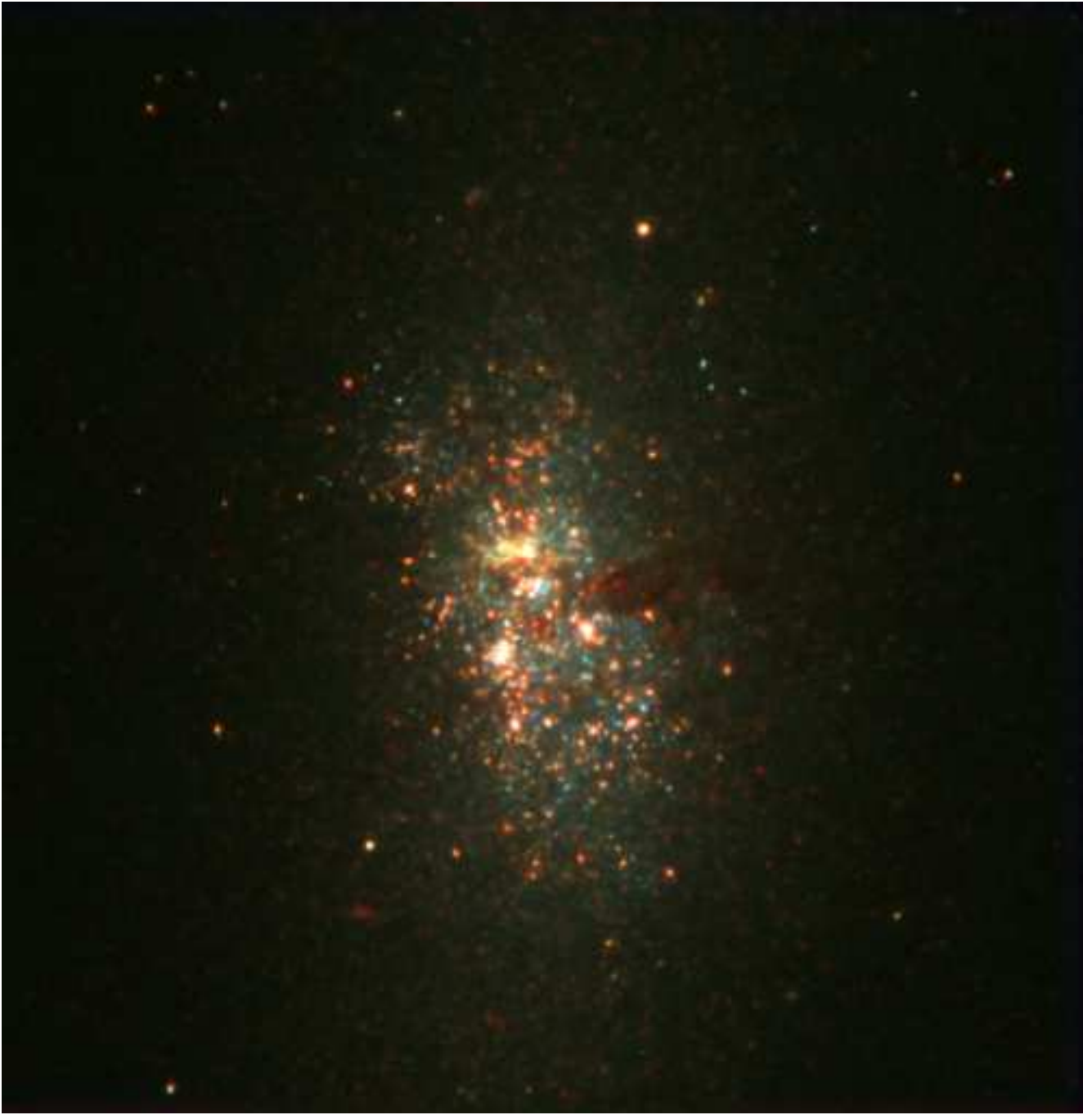}
    \caption{Three-color image of NGC 5253 in the V band (blue), I band
    (green) and Ks band (red). North is up and east is right. The image is
    $80\arcsec \times 80\arcsec$ in size.}
    \label{tricromia}
\end{figure*}
The absolute photometric calibration of the Ks band image was performed
with photometric standard stars, while for WFPC2 images the zero points listed in the
\textit{HST Data Handbook} (Baggett et al. \cite{baggett}) were used.
The transformation from the WFPC2 photometric system to UBVRI was performed according to
Holtzman et al. (\cite{holtzman}), who supply a relation between the HST STMAG
system magnitude and the color (V-I) of the target. That relation was inverted to
derive the color (V-I) as a function of the observed (F547M-F814W). The coefficients,
which are also functions of the color, were chosen using as initial guess the (V-I) color
derived with the \textsf{synphot} package (Simon \& Shaw \cite{simon}) for a B5
star with the same (F547M-F814W). The corrections for the magnitudes were derived
from those on the color assuming $(V-F547M)=-0.036$. It must be noted that the
difference between the F814W and the I filter is more than 1 mag, leading
to a large color term ($\sim 1.2$ mag).

The narrow band images in the F656N and F673N filters were calibrated using the
PHOTFLAM keyword in the image header.

\section{Photometry of the clusters and the color-color diagram} \label{phot}

In Fig. \ref{tricromia} we present a three-color image of NGC 5253 obtained by
the combination of HST-F547M (blue), HST-F814W (green) and VLT-Ks (red).
If we assume as a first and crude approximation that the red objects in the
image are, on average, more extincted than the others, a simple examination of
this image tells us that the dust extinction is quite inhomogeneous across the
galaxy.
Regions of blue and red objects, in fact, appear to be randomly distributed and,
besides the well known dust lane, there is no evidence of a regular pattern or 
structure.
A number of likely highly extincted red objects are visible only in the Ks band.
A more quantitative idea can be obtained from the photometry of the sources
and from the analysis of their color-color diagram.\\
To extract the photometry of the sources we have tested two different methods.
In the first one the background was estimated in an annulus of $0\farcs5$ around each
object, using a constant circular aperture of $0\farcs5$ for photometry. This
technique provides reliable results only for the brightest and isolated objects
(see C97), while the error obtained from a sample of artificial sources of fixed
magnitude $m_V=18.4$ is up to about 0.5 mag where the sources are particularly 
crowded.
To improve the photometry of the faint sources we have chosen a different method to
estimate and subtract the background which consisted in fitting the luminosity profile
of the underlying galaxy.
All the sources with more than $2.5\ \sigma$ detection were rejected to
compute the fit, and Legendre polynomials of order 40, 30 and 28 were used
respectively for the V, I and Ks band image. The orders chosen were the ones that
gave the best fit and minimized the residuals. The obtained fitted surface was
then subtracted from the image of the galaxy, and the resulting image was used for
the photometry. The error for the photometry for the same sample of artificial objects
is reduced to about 0.05 mag.\\
\begin{figure*}
    \begin{center}
    \includegraphics[width=0.95\textwidth]{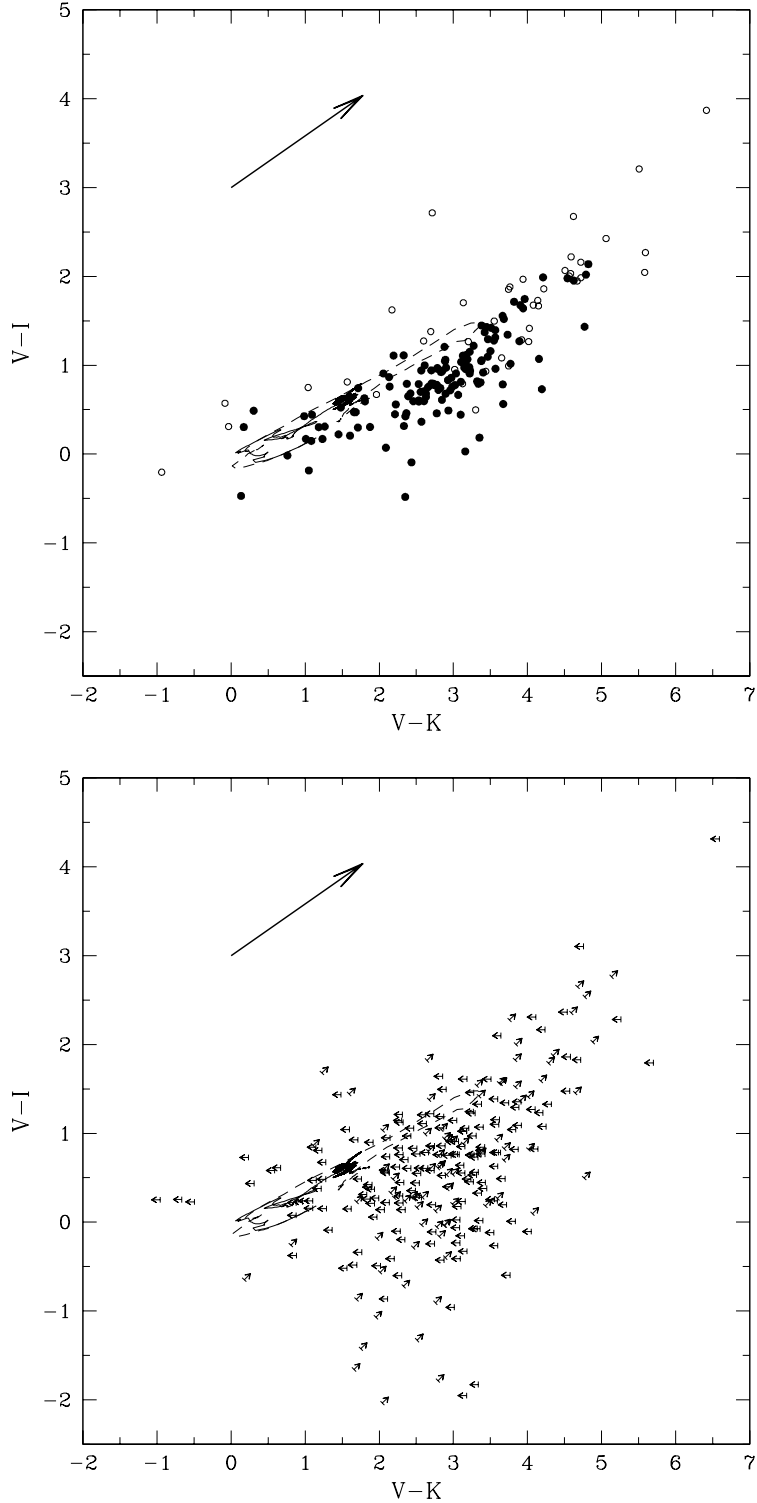}   
    \caption{Color-Color plot of the observed clusters in NGC 5253.
    \textit{Upper panel:} the circles are the clusters detected in all all bands 
    (V, I, and Ks), the filled ones with a mean of the uncertainties in the two
    colors of less than 0.2 mag and the open ones with errors between 0.2 and 0.44
    mag. The solid line is the SB99 model of an instantaneous burst
    of star formation, with $Z=0.004$, assuming a Salpeter IMF with
    $\alpha=2.35$ and $M_{\textrm{up}}=100\ M_{\sun}$, while the dot-dashed line is the
    same model but with $Z=Z_{\sun}$, where the red supergiants loop is evident
    (see text). The arrow in the upper left corner is the screen extinction
    vector for $A_V=2$ mag.
    \textit{Lower panel:} the arrows are upper and lower limits on the colors
    for the objects detected in V and I but not in Ks (the ones pointing to the
    left) or in I and Ks but not in V (pointing to the upper right).
    The SB99 models and the screen extinction vector are plotted as in the upper
    panel.}
    \label{colcol}
    \end{center}
\end{figure*}
The detection of the sources was performed using SExtractor (Bertin \& Arnouts
\cite{bertin}) with parameters set to have secure detections, $5\ \sigma$ per pixel
over an area of 4 pixels. The V and I band images have been previously smoothed,
by convolving them with a Gaussian, to reduce them to the angular resolution of the
Ks band image (0.4 arcsec).
In this way it was possible to match multiple sources, resolved only in the
HST frames, with a single unresolved source in the ISAAC frame.
In the V band smoothed image 491 objects were recovered, 524 in the smoothed I band
one, and 357 in the area of the Ks band frame covered by the HST observations. \\
A geometric coordinate transformation between the different images of the
galaxy has been computed using as reference 50 bright sources present in all the
frames.
The transformation was then used to match sources in the three different images.
We consider a source as matched in two of the frames when the distance between
the centers in the different images, referred to the same coordinate system, is
less than 3 pixels (or $0\farcs44$).
With these constraints, 375 sources were matched between the smoothed V and I band
frames, and 192 between the Ks band and the smoothed V band ones. 178 sources were 
matched in all the three bands. For all other
objects, and for the sources matched only in two of the filters, we give an
upper or lower limit for the colors, using the $80\%$ completeness magnitudes for a
$5\ \sigma$ detection in the frame where the objects are not recovered. This is about
21 for the V and I band and 18 for the Ks band.

The photometry of the detected sources was carried out on the background-subtracted
images, using a circular aperture of $0\farcs75$, 1.87 times the PSF, i.e. 5
pixels in the Ks band image and 7.5 pixels in the smoothed V and I band ones.

One of the main sources of error for the photometry is the crowding of 
the sources. To estimate this contribution we used the sample of clusters 
detected in the unsmoothed images in each band to calculate the probability P(L) 
of finding an object of  a given luminosity L in the aperture used for photometry, 
and then we summed all the contributions L $\cdot$ P(L). 
We ended with photometric errors due to crowding of about 0.12, 0.11, 0.04 mag 
for the V, I and K band respectively. 

The resulting V-I/V-K color-color plot is shown in Fig. \ref{colcol}.
The uncertainties in both colors vary between 0.12 to 0.44 mag.
The circles in the upper panel are the sources matched in all three bands, 
the filled ones with a mean of the uncertainties in the two colors of less than 0.2 mag and 
the open ones with errors between 0.2 and 0.44 mag.
The arrows in the lower panel are upper and lower limits on the colors for the objects
detected in V and I but not in Ks (the ones pointing to the left)
or in I and Ks  but not in V (pointing to the upper right).
The screen extinction vector for $A_V=2$ is also reported. The solid line
is the prediction of the Starburst99 model (SB99, Leitherer et al. \cite{leitherer}) for an
instantaneous burst with metallicity $Z=1/5 Z_{\odot}$, assuming a Salpeter IMF with
$\alpha=2.35$, $M_{\textrm{low}}=1\ M_{\sun}$ and $M_{\textrm{up}}=100\ M_{\sun}$, with an age coverage
$10^5-10^9$ yr, while the dot-dashed line is the same model but with $Z=Z_{\sun}$. \\

\section{Analysis and Discussion} \label{discuss}

From a comparison of the data with the models it is evident that a large number of
sources is affected by a significant red excess. The degeneracy between age and
extinction, both producing a reddening of the clusters, prevents us from determining
these two quantities independently using the color-color diagram alone.
This degeneracy will be partly removed in Sect. \ref{ages}, and the derived ages
will be used to evaluate the extinction toward the clusters in Sect.
\ref{extinction}. This will enable us to build the luminosity and finally the
mass function in Sect. \ref{lumfunc} and \ref{mass} respectively.\\
But first of all, note that there is a significant difference
between the solar and sub-solar models.
The first extends much further in the red mainly because of the  red supergiant phase
between 5 and 15 Myr of age, this phase being artificially suppressed in the sub-solar
model. In fact the sub-solar models fail to reproduce the right number of red
supergiants, and there are indeed indications that their number increases rather than 
decreases as the metallicity is reduced (e.g. Maeder \& Meynet \cite{maeder},
Eggenberger, Meynet \& Maeder \cite{eggenberger} and references therein). Moreover,
the duration of the red phase is expected to increase as the abundance decreases
(Maeder \& Meynet \cite{maeder}).
In other words, neither model can be considered a perfect reference, as
this effect has important consequences for the predicted colors (Origlia et al.
\cite{origlia}). In the following we will consider both models, comparing the
results obtained in the two cases. \\

\subsection{Star contamination} \label{stars}

Since most of the sources detected are barely resolved or not at all, our
sample could be significantly contaminated by stars, both
foreground and galaxy members. Foreground stars should not be a
source of confusion, as the standard Milky Way star count models
(e.g. Ratnatunga \& Bachall, \cite{ratnatunga}) predict roughly
2-3 foreground stars in our field of view with $m_V < 21$.
However, individual blue and red supergiant stars can have
absolute magnitudes as bright as $M_V \simeq -9$ (Humphreys
\cite{humphreys}), that is $m_V \simeq 18.6$ at the distance of
NGC 5253. As a consequence, our cluster sample can be contaminated
by single bright stars. \\
As only few of the sources in NGC5253 are brighter than $M_V < -9$, 
we used the high angular resolution provided by the WFPC2
images to distinguish the resolved clusters from the point like
objects. The archive WFPC2 images were centered on the WF3 chip,
which provides a PSF with FWHM $\sim 0\farcs13$, corresponding 
to $\sim 2$ pc at the distance of the galaxy. This is slightly
smaller than the typical dimensions of young star clusters, so
that also in the WF images clusters are in general only barely
resolved. The FWHM of the sources matched in the three bands was
measured in the F547M original image and compared with the PSF 
provided by TinyTim for the same configuration 
as well as with the PSF 
of bona-fide stars in the field. Objects whose PSF was indistinguishable 
from point sources, i.e. with $FWHM < 0\farcs153$, were then discarded from 
the following analysis. This way we removed from the sample 
$\sim 35\%$ of the 178 sources matched in all the filters, 
ending up with 115 clusters.

\subsection{The ages of the clusters} \label{ages}

\begin{figure}
    \centering
    \resizebox{\hsize}{!}{\includegraphics{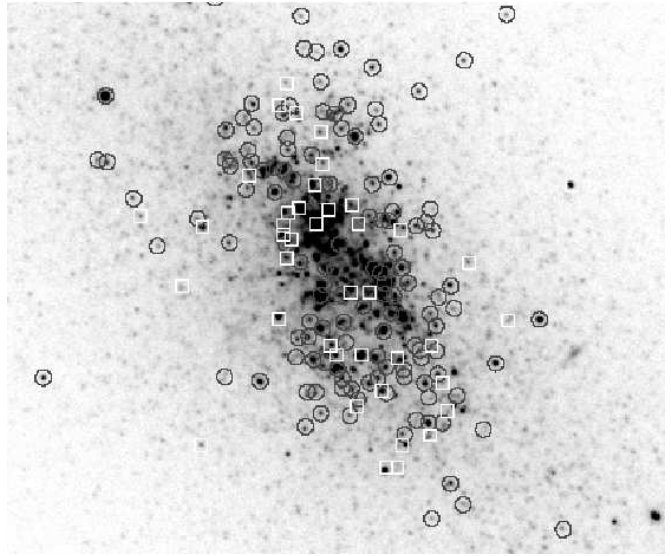}}
    \caption{Spatial distribution of older (gray circles, age $>$ 20 Myr) and
    younger (white rectangles, age $<$ 20 Myr) matched clusters, compared with the
    VLT Ks band image.}
    \label{agedistrib}
\end{figure}

As already mentioned, the ages of the clusters cannot be derived in
a unique way from the colors, so that it is essential to have an
independent age indicator. The Equivalent Width (EW) of the
H$\alpha$ emission line can be used for this purpose, as it
decreases monotonically with age, according to the
SB99 model, with little dependence on metallicity. Therefore we
used this method to derive the ages of 51 clusters matched in V, I
and Ks which have also been detected in H$\alpha$; the ages are obtained
by comparison of the measured EW with the SB99 model prediction.
The derived ages span a range between 3 and 19 Myr. \\
According to the model, older clusters are not detectable in the
H$\alpha$ image because of their faint emission, since very few O and
B stars are present to ionize the gas. To check that young faint
objects, with low flux but large EW, were not placed in the
sample of old objects, we derive an estimate of the flux expected
in the F656N (H$\alpha$) filter for all the objects in the sample.
The flux in the H$\alpha$ continuum, F673N filter, was evaluated
for the objects matched in the V and I bands by linear
interpolation between the fluxes measured in these two bands. In
this way it is possible to calculate the expected flux in
H$\alpha$ for all the objects as a function of the EW. If we
assume an EW $>45\ \AA$, corresponding to an age of 10 Myr according to
the SB99 model, the derived H$\alpha$ flux is above our detection
limit for all the objects, so that we can state that we are not
missing any source with EW greater than $45\ \AA$ and younger than 10
Myr. Therefore our age estimation provides reliable values for
``young" clusters detected in H$\alpha$, and a lower limit of 10
Myr for the ``old" ones, yielding a division of the sources in
two subsections. In Fig. \ref{agedistrib} we plot the spatial
distribution of ``old" (grey circles) and "young" (white squares) clusters.  
No clear correlation is present between the spatial distribution of the clusters
in the galaxy and their ages, although there is a group of young
objects located at the position of cluster-5. \\
With this constraint on the age, it is possible to derive the
extinction by the color excess once the SB99 model is assumed, as
explained in the following section.

\subsection{Extinction and dust} \label{extinction}

We derived the extinction using our color-color diagram and
the color excess of the clusters with respect to the SB99 model,
once the age of each source is known. 
We assume that the discrepancy between the model and the data
points is produced by the effect of screen extinction and emission
by dust. If we assume a dust temperature of 1000 K, its emission
contributes significantly only in the K band, and as a consequence
moves the points to the right, increasing V-K. Combining these two
effects, it is possible to derive the extinction for the clusters
of known age, for which we know the theoretical position in the
diagram. Visual screen extinctions in the range 0 to 7.5
magnitudes were found, with an average of 1.6 mag. An estimate of
the extinction for the ``old" clusters can be obtained assuming 
a reference age of 20 Myr for all these objects. This is obviously 
a crude approximation; however, the change of colors with age is much
smaller for clusters older than 20 Myr ($\Delta (V-I) \sim
0.2$ between 20 Myr and 1 Gyr) than for younger ages ($\Delta
(V-I) \sim 1.9$ between 0 and 20 Myr). This way we obtain
extinctions in the same range as for the young clusters.

Note that in using the color diagram to derive the
screen extinction we are underestimating the actual column density
of obscuring material, as we are neglecting the scattering by dust
grains, which strongly affects colors. In fact, while absorption reddens 
the intrinsic spectrum, scattering makes it bluer by reflecting some light 
in our direction (e.g. Witt et al. \cite{witt}). The net result is that 
the magnitude of the extinction decreases for a given optical depth when 
scattering is taken into account. In particular, Vanzi \& Sauvage (\cite{vanzi}) 
find that the extinction is 3-4
times higher than in the simple screen case when the effect
of scattering is included for the extreme case of cluster 5. \\
The X-ray and radio data available in the literature could not be used to 
measure the extinction, as none of the objects detected at those wavelengths
can be clearly associated with optical or NIR sources (Turner et al. \cite{turner}, Strickland \& 
Stevens \cite{strickland}, Summers et al. \cite{summers}). \\

\subsection{The Luminosity Function} \label{lumfunc}

\begin{figure}
    \centering
    \resizebox{\hsize}{!}{\includegraphics{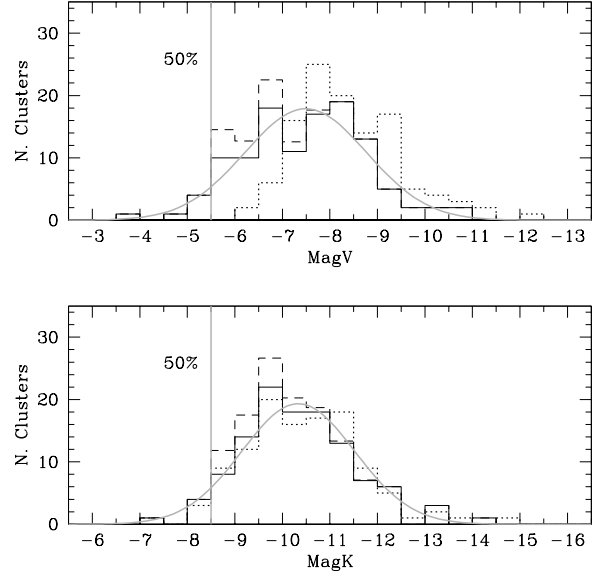}}
    \caption{The luminosity function of the sources observed in V and Ks band.
    The solid line is the observed function, and the dotted one is after
    correction for extinction. The dashed line is the completeness-corrected
    function, where the correction is applied to the 50\% completeness limit.
    The Gaussian curves are the best fit to the observed distributions.}
    \label{funclum}
\end{figure}

In Fig. \ref{funclum} we show the V and Ks band luminosity functions of the 
selected sources. The solid line is the observed function, while the dotted one is after
correction for extinction. As expected, the effect of the extinction is much more
relevant in V than it is in Ks.\\
Completeness tests were performed in, for both the V band and Ks band
images adding in each frame 80 artificial objects, concentrated in
the central part of the galaxy, with shapes derived by using
isolated, high S/N clusters from the same image. The sources were
added each time with different magnitudes, and recovered with the
same technique used for the real clusters. The catalogue of the
recovered sources was then cross-correlated with the catalogue of
the added artificial sources to derive the fraction of recovered
objects for each luminosity channel, and thus the completeness
correction. The dashed line in Fig. \ref{funclum} is the
luminosity function with the completeness correction obtained from
these simulations. The correction is applied to the 50\%
completeness limit magnitude, where half of the artificial objects
were recovered. The two corrections, for extinction and for completeness, 
cannot be combined to produce the intrinsic luminosity function. 
In addition, the luminosity function was not corrected for age effects, 
as we have only a lower limit for clusters not detected in H$\alpha$. 

The luminosity functions show a peak about 1.5 mag brighter than our 50\%
completeness limit, both in the V ($M_V \left(\textrm{\small peak}\right)=-7.5$) 
and the Ks band ($M_K \left(\textrm{\small peak}\right)=-10.3$). The significance 
of this peak is highest in the K band where the effect of the extinction 
is minimum.
The luminosity distribution $\psi(L)$ is remarkably close to the shape of the old
globular cluster luminosity function (roughly Gaussian with the peak or turnover
magnitude at $M_V \simeq -7.4$, e.g. de Grijs et al. \cite{degrijsb}), but quite
different from the power law shape ($\psi(L)\ Tl \propto L^{-\alpha}\ dL$) 
derived by Whitmore et al. (\cite{whitmore}) for the luminosity function of young 
(age $<$ 100 Myr) clusters in the Antennae galaxy.

A systematic underestimate of the completeness correction would be needed to obtain
a power law from the observed distribution. This cannot be completely
ruled out, especially below the 80\% completeness limit, which corresponds to the
detected peak.

\subsection{The Mass Function}  \label{mass}

With all the information gathered about the star cluster
population in NGC~5253, it is possible to provide, under a few
assumptions, an estimate of the mass function of these objects. 
To derive the mass we will use here the Ks luminosity as it is less
affected by the extinction and is a better tracer of the mass.\\
To convert the Ks luminosity into mass we have to assume a Ks-to-mass 
ratio. Vanzi \& Sauvage (\cite{vanzi}) have studied in detail
the spectral energy distribution (SED) of Cluster-5 using all data
available in the literature, and developed a model of both the
stellar cluster and the surrounding dust shell. Here we assume that the
shell geometry of cluster and dust, the dust properties and the
stellar SED of Cluster-5 are typical of the sources in NGC~5253.
Under this assumption, we have run a set of models at different ages 
and with different amounts of optical opacity, and therefore of dust, in 
order to obtain the light-to-mass ratio, or the ratio between the K luminosity
and the total mass. 
The model tabulates the K luminosity to cluster mass ratio for a range of 
ages and extinction. For each observed cluster we use our determined age 
and extinction to select the most appropriate model and use the corresponding 
light-to-mass ratio to derive the cluster mass. 
\begin{figure}
    \centering
    \resizebox{\hsize}{!}{\includegraphics{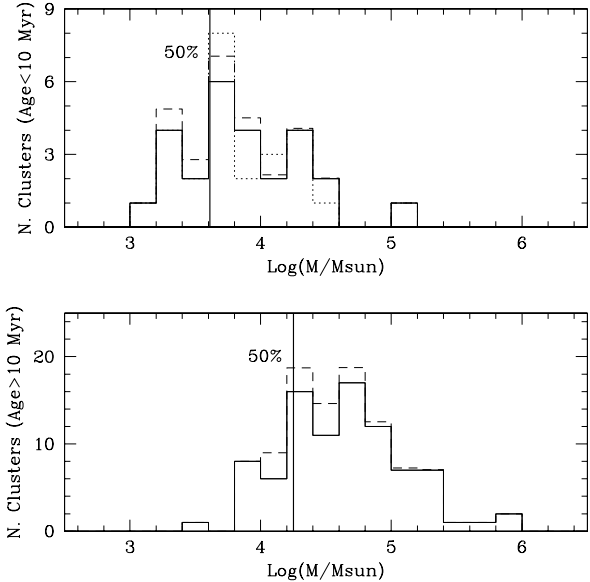}}
\caption{Mass function of the young clusters detected in H$\alpha$ (upper panel) 
and of the older ones (lower panel). The 50\% completeness limits are marked by 
vertical lines. The dotted line in the upper panel uses the solar model as reference, 
while the dashed lines are the mass functions after the completeness correction has been 
applied.}
    \label{massfunc}
\end{figure}

This way it is possible to convert the luminosity function into
a mass function. We have done this using both the solar (dotted)
and sub-solar (solid) models to derive the screen extinction. The
results are shown in Fig.~\ref{massfunc}. The upper panel shows
the mass function of the sources younger than 20 My detected in
H$\alpha$ obtained using the solar and sub-solar models as
reference. The difference between the two is not
significant, as few of the clusters have an age corresponding to
the supergiant phase, where the loop distinguishing the two models
is located. The lower panel shows the mass function of the
clusters not detected in H$\alpha$ and therefore older than 10 Myr, 
assuming a reference age of 20 Myr in this case. 
The dashed lines are the mass functions
after the completeness correction has been applied. The same number and
luminosity of missed clusters derived from the correction of the
luminosity function was used to compute the completeness
correction for the mass function. Their mass was calculated from
the luminosity assuming the average extinction of the observed
clusters ($A_V=0.98$ mag), the average age of 8 Myr of the young
clusters and a reference age of 20 Myr for the older ones.

In the distribution of the old objects a turnover is detectable 
at about $5 \cdot 10^4 M_{\odot}$, 2.5 times more massive than the 
50\% completeness limit. In the distribution of the young objects no peak 
is observable above the completeness limit, and the data are well 
reproduced by a power law. The best fitting power law index for the mass
function $\psi(M)\ dM \propto M^{-\beta}\ dM$ of the young objects is
$\beta = 1.56$, while $\beta=1.82$ is obtained 
for objects more massive than the peak in the old cluster
distribution.

Using a selection criterion between clusters and single bright stars 
based only on the TinyTim model PSF for the V band image, instead of 
using stars in the field, we discard as unresolved sources around 15\% 
of the objects matched in all the three bands. 
In this case we obtain a slightly larger exponent for the 
young cluster distribution, $\beta = 1.71$, while the peak in the 
luminosity and and mass functions of old clusters is still present.
The power law index $\beta$ derived for the mass function 
of younger clusters is in the same range as what has been found in a 
wide variety of galaxies (e.g. de Grijs et al. \cite{degrijsd}), 
and fully consistent with the mass function index of clumps in a 
large number of molecular clouds (see e.g. Kramer et al. \cite{kramer}
and references therein).\\
The currently most popular model for the dynamical evolution of
star clusters predicts that the power law mass and luminosity
functions of young star clusters, detected in many nearby galaxies
(see Whitmore et al. \cite{whitmore} and Zhang \& Fall,
\cite{zhang} for the ''Antennae'' galaxies), will be rapidly
transformed into the Gaussian functions of old globular clusters,
under the action of several effects (e.g. Fall \& Zhang
\cite{fall}). These include the preferential depletion of low-mass
clusters, by evaporation due to two-body relaxation and by tidal
interactions with the gravitational field of the host galaxy, and
the preferential disruption of high-mass clusters by dynamical
friction. Till now there was only one example of a turnover in the
mass and luminosity distribution for intermediate-age ($\sim 1$
Gyr) clusters, reported by de Grijs , Bastian, \& Lamers
(\cite{degrijsc}) in the center of M82. 

Our data may be consistent with a scenario where young, power-law-distributed 
clusters are superimposed on an older population with
the same initial power law index that has already developed a
turnover in the luminosity and mass distributions. Comparing the
total mass of the youngest ($4.1 \cdot 10^5\ M_{\odot}$) 
to the total mass of the oldest clusters ($7.4 \cdot 10^6\ M_{\odot}$), we find 
that the most recent burst of star formation is at least $\sim 20$
times less intense than the previous ones.\\
The total mass of the matched clusters is $7.8 \cdot 10^6\
M_{\odot}$, which has to be compared with the total mass of the
galaxy of $\sim 6 \cdot 10^9\ M_{\odot}$ (Vanzi \& Sauvage
\cite{vanzi}).

\section{Summary and conclusions}

The conclusions of our work can be summarized as follows.
\begin{enumerate}
\item We have detected a large number of compact sources in the starburst
galaxy NGC~5253 and built their V-I/V-Ks color-color diagram. From the comparison of
the data points with models we found that most sources are affected by a significant
color excess, which we mostly ascribe to dust.
\item 178 sources were matched in all the three bands V, I, and K.
To take into account the stellar contamination by single bright stars 
we have discarded on the basis of their point-like profiles around $35\%$ 
of these sources, ending up with 115 clusters.
\item We have partially broken the age-extinction degeneracy of the color-color
diagram and derived the luminosity and mass functions of the clusters.
\item The ages of the clusters, as measured from their H$\alpha$ emission, span a
range from few to 19 Myr. Many of the clusters are not detected in H$\alpha$,
implying ages $>$ 10 Myr.
\item  The clusters younger than 20 Myr show a mass function that can be modeled with
a power law $\psi(M) \propto M^{-\beta}$, with an exponent $\beta = 1.6$. For  
objects more massive than $5 \cdot 10^4 M_{\odot}$ in the older cluster distribution 
a value of $\beta=1.8$ is obtained. These values are in the same range as what  
has been found in a wide variety of galaxies, including interacting galaxies 
and more quiescent objects (de Grijs et al. \cite{degrijsd}), and fully consistent 
with the mass function index of clumps in a large number of molecular clouds
 (Kramer et al. \cite{kramer}).
\item Both the luminosity functions and the mass function of the oldest clusters show a
peak or turnover above our 50\% completeness limit. The luminosity functions
in the V and Ks band, in particular, show a Gaussian profile peaked at $M_V=-7.5$ and
$M_K=-10.3$. The shape of the distribution is remarkably close to that of the old globular
clusters, roughly Gaussian with a peak at $M_V \simeq -7.4$, and quite
different from the power law distribution found in other young cluster populations.
If the peak in our data is confirmed, this galaxy may provide a second example
of a turnover in an intermediate age cluster population maybe even younger than in
M82.
\end{enumerate}

\begin{acknowledgements}
We want to thank the anonymous referee for useful comments and suggestions.
We are also grateful to Ian Stevens for providing us with results of his work before
publication. GC acknowledges support from the European Southern Observatory
during his visit to Chile.
\end{acknowledgements}


\end{document}